\documentclass[aip,apl, reprint,superscriptaddress, twocolumn]{revtex4}
\usepackage{graphicx}
\usepackage{dcolumn}
\usepackage{bm}

\usepackage{color,ulem}

\begin{document}
\title{Ultrafast plasmonics using transparent conductive oxide hybrids in the epsilon near-zero regime.}
\author{Daniel Traviss}
\affiliation{Physics and Astronomy, Faculty of Physical and
Applied Sciences, University of Southampton, Highfield,
Southampton SO17, 1BJ, United Kingdom}
\author{Roman Bruck}
\affiliation{Physics and Astronomy, Faculty of Physical and
Applied Sciences, University of Southampton, Highfield,
Southampton SO17, 1BJ, United Kingdom}
\author{Ben Mills}
\affiliation{Optoelectronics Research Centre, Faculty of Physical
and Applied Sciences, University of Southampton, Highfield,
Southampton SO17, 1BJ, United Kingdom}
\author{Martina Abb}
\affiliation{Physics and Astronomy, Faculty of Physical and
Applied Sciences, University of Southampton, Highfield,
Southampton SO17, 1BJ, United Kingdom}
\author{Otto L. Muskens}
\affiliation{Physics and Astronomy, Faculty of Physical and
Applied Sciences, University of Southampton, Highfield,
Southampton SO17, 1BJ, United
Kingdom}\email{O.Muskens@soton.ac.uk}

\date{\today}

\begin{abstract}
The dielectric response of transparent conductive oxides near the bulk plasmon frequency is
characterized by a refractive index less than vacuum. In analogy with x-ray optics, it is shown that this regime results in total external reflection and air-guiding of light. In addition, the strong reduction of the wavevector in the ITO below that of free space enables a new surface plasmon polariton mode which can be excited without requiring a prism or grating coupler. Ultrafast control of the surface plasmon polariton mode is achieved with a modulation amplitude reaching 20\%.
\end{abstract}

\maketitle

Epsilon-Near Zero (ENZ) materials are a class of optical materials
characterized by a real part of the dielectric function close to
zero. ENZ materials are of interest for a range of applications
including tailoring of directional emission and radiation phase
patterns,\cite{Enoch02, Alu07,Zayats09} air-guiding of
electromagnetic waves,\cite{Schwartz04} and electromagnetic
tunnelling devices.\cite{Silveirinha06,Vesseur12,Lu12} While a lot
of effort is aimed at achieving an ENZ response using artificial
metamaterial resonators, some naturally occurring materials also
show a strong reduction of the permittivity below that of vacuum.
An example of naturally occurring low-index materials are noble
metals where the optical permittivity $\epsilon$ is governed by
the collective motions of the free electron gas known as bulk
plasmons. According to the Drude model, the permittivity is given
by
\begin{equation}
\epsilon(\omega)=\epsilon_{\infty} - \frac{\omega_{\rm pl}^2}{\omega^2+i\omega \gamma} \, ,
\end{equation}
where $\gamma$ denotes the damping rate of the free electrons and
the plasma frequency $\omega_{\rm pl}$ is given by $\omega_{\rm
pl}=(N e^2/\epsilon_0 m)^{1/2}$. Around the (screened) bulk
plasmon frequency $\omega_{\rm bp} \equiv \omega_{\rm pl}/\sqrt
\epsilon_\infty$, the real part of the permittivity shows a
transition from negative to positive values. Noble metals have
carrier densities $N$ exceeding $10^{22}$~cm$^{-3}$, therefore
their bulk plasmon plasma frequency is located in the UV region.
In contrast, highly doped semiconductors typically have electron
densities below $10^{19}$~cm$^{-3}$ and can be well described by
the Drude model with a bulk plasmon plasma frequency in the THz
range. Transparent conducting oxides (TCOs), with an electron
density inbetween that of bulk metals and doped semiconductors,
show a bulk plasmon frequency in the near-infrared. The resulting
combination of a metal-like response in the infrared and a
dielectric optical response in the visible region has stimulated
application of TCOs as transparent electrical contacts and as heat
reflecting windows. Recently, metal oxides such as indium-tin
oxide (ITO) and aluminium-doped zinc oxide have received interest
for their plasmonic response in relation to metamaterials and
transformation optics.\cite{Boltasseva,Noginov11} The plasma
frequency can be tuned by controlling the electron density using
electrical or optical methods, opening up opportunities for
near-infrared electro-optic or optical modulators
\cite{Lu12,Feigenbaum10, Abb11} and sensing
devices.\cite{Franzen02,Odom11}

Pioneering studies by Franzen and co-workers have investigated the
plasmonic response of ITO and ITO-gold hybrid structures in the
metallic (negative epsilon) regime of ITO.\cite{Franzen02} Next to
a conventional surface plasmon polariton mode for thick ITO films,
a new polariton mode associated with the bulk plasmon resonance
was observed for 30~nm thin films. By using hybrid ITO-gold
layers, the balance between the two types of plasmonic modes could
be controlled. In analogy with metal nanoantennas, ITO and AZO
colloidal nanoparticles and nanorods revealed pronounced surface
plasmon resonances in the infrared
range.\cite{Kanehara09,Odom11,Buonsanti11}

\begin{figure*}[t]
\includegraphics[width=15.0cm]{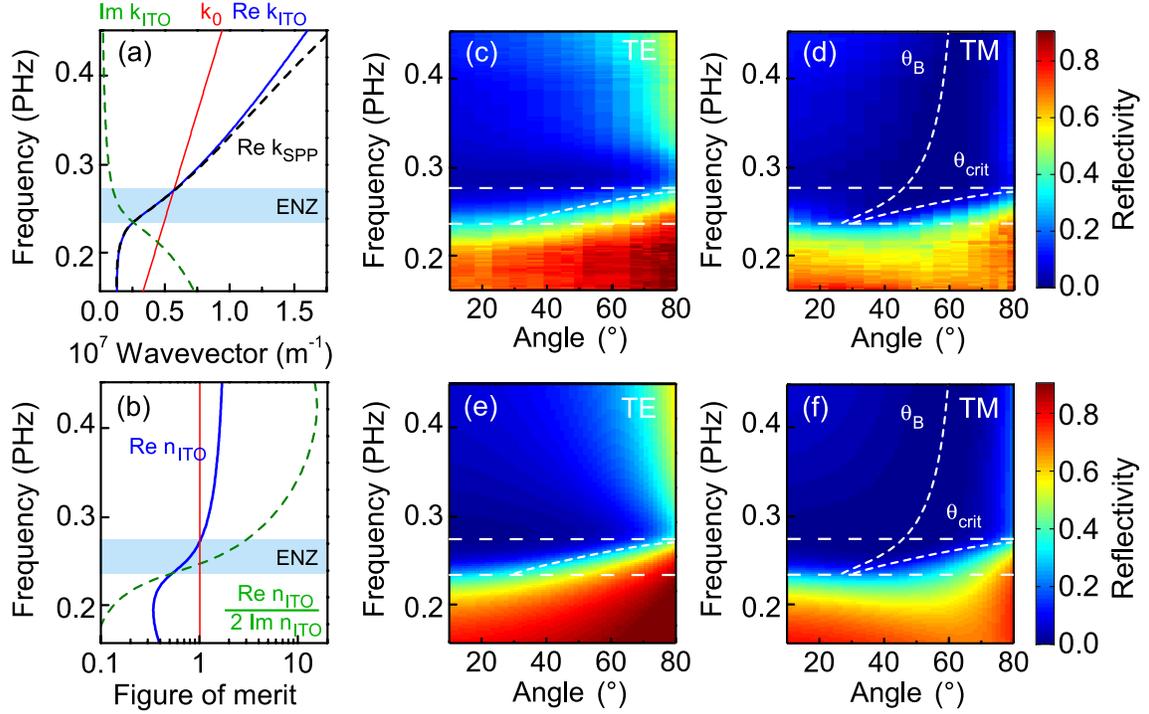}
\caption{\label{fig:dispersion} (Color online) (a) Photon dispersion relations
for air (red), ITO (blue) and the ITO-Au SPP (dash, black), with
imaginary part for ITO (short dash, green). (b) Same for real
refractive index ${\rm Re} \, n_{\rm ITO}$ and the propagation length in number of wavelengths, ${\rm Re} \, n_{\rm ITO}/2 {\rm Im} \, n_{\rm ITO}$. Reflectivity as a
function of angle, with (white line) calculated TERF critical
angle. Blue shaded area in (a) and dotted lines in (b) indicate
epsilon-near zero (ENZ) regime. }
\end{figure*}

Here, we explore the intermediate regime where ITO shows a
dielectric response with a refractive index below that of vacuum.
In this studies, we used commercial ITO slides (Lumtec) of 350-nm
thickness with a sheet resistance of $5\pm 1$~$\Omega/$square.
Figure~\ref{fig:dispersion}(a) shows the calculated dispersion
relation for these ITO samples (blue line) using the Drude model
with experimentally obtained parameters $m=0.4 m_e$,
$\gamma_D=0.2$~PHz \cite{Franzen02} and $N=1.13 \times
10^{21}$~cm$^{-3}$. The thin line (red) indicates the dispersion
relation of light in vacuum. A low-index regime is observed in the
frequency range between $0.24 - 0.27$~PHz, which characterized by
a superluminal phase velocity and a correspondingly reduced
wavevector. The dielectric ENZ regime covers the range of
(complex) wavevectors given by the condition ${\rm Im}\, k_{\rm
ITO}<{\rm Re}\, k_{\rm ITO}<k_0$, as indicated by the shaded area
in the dispersion relation. Figures of merit for the ENZ response
are the refractive index ${\rm Re} \, n_{\rm ITO}$ and the
propagation length in units of wavelength ${\rm Re} \,  n_{\rm
ITO}/2 {\rm Im} \, n_{\rm ITO}$, as shown in
Fig.~\ref{fig:dispersion}(b). While the refractive index increases
from 0.48 to 1.0 over the ENZ window, the propagation length of
light in the ITO layer increases from 0.5 to 3.2 wavelengths.

Light incident on an interface between two media from the medium
with the higher refractive index undergoes total internal
reflection above a critical angle $\theta_{\rm crit}$. Similarly,
at the interface between air and an ENZ medium, a similar effect
of total external reflection occurs. Total external reflection at
grazing angles is a well-known phenomenon in x-ray optics, where
it is commonly used in reflective focusing optics, waveguides, and
surface analysis.\cite{Aschenbach85}
Figures~\ref{fig:dispersion}(c,d) show the reflectivity of the
ITO-glass slide for various angles of incidence for TM and TE
polarizations. The dashed lines indicate the estimated positions
of the critical and Brewster's angles $\theta_{\rm crit}$ and
$\theta_{\rm B}$ in the absence of absorption, i.e. using
$\theta_{\rm crit}=\sin^{-1} ({\rm Re} \, n_{\rm ITO})$ and
$\theta_{B}=\tan^{-1} ({\rm Re} \, n_{\rm ITO})$. The critical
angle ranges between $30^\circ$ and 90$^\circ$. For both
polarizations, an increase in reflectivity is observed in the ENZ
regime above the critical angle. The exact definitions for
$\theta_{\rm crit}$ and $\theta_{\rm B}$ loose validity in the
presence of absorption, and for strong absorption they tend to
shift toward larger angles.\cite{Meeten97} To compare the exact
behavior we calculated Fresnel's reflectivities using the Drude
model with optimized free carrier density of $N=1.13\times
10^{21}$~cm$^{-3}$. The resulting Figs.~\ref{fig:dispersion}(e,f)
show good agreement with the experimental data and confirm the
presence of total external reflection in the NEZ regime of ITO.

In order to assess the air-guiding effect associated with total
external reflection in ITO, we changed the configuration to a
planar waveguide by aligning two ITO-slides with a tunable
separation of several hundred $\mu$m. The optical beam was focused
to a spot of 50 $\mu$m in diameter using a focusing lens with
numerical aperture of 0.02. By tuning the space between the
slides, we achieved up to 8 multiple reflections. The resulting
transmission spectrum through the waveguide is shown in
Fig.~\ref{fig:dispersion} for different angles of incidence. We
identify three regimes, respectively at energies below, in, and
above the ENZ regime. Below the ENZ window, the metallic
reflectivity of ITO results in a weakly varying transmission of
several percent. This transmission is low because of insertion
losses as well as because of the limited single-pass reflectivity
of up to 75\% giving rise to a multi-pass transmission below 10\%.
Above the ENZ regime, the transmission contains contributions from
the air-ITO and the ITO-glass interface, which are overall
relatively small but increasing toward large angles of incidence.
In the ENZ regime, large variations of the transmission are found
when changing the angle of incidence by only a few degrees. A
pronounced spectral dip is observed for TM polarization at angles
below 80$^\circ$, which is caused by a combination of reduced
reflection of the ITO-air interface and increased absorption in
the ITO layer. Above the critical angle, the waveguide
transmission dramatically increases by more than three orders of
magnitude around $0.25$~PHz. These results indicate a potential
application of TCO air-guided waves using total external
reflection, for example in optical sensors or modulators.

\begin{figure}[t]
\includegraphics[width=8.5cm]{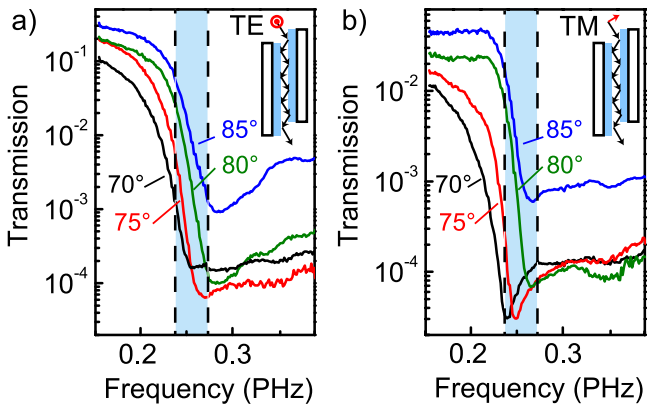}
\caption{\label{fig:waveguide} (Color online) Transmission of
air-guided ITO-cladded waveguide for 8 multiple reflections, for
TE (a) and TM (b) polarizations and for angles of incidence
between 70$^\circ$ and 85$^\circ$. Blue shaded areas indicate
epsilon-near zero (ENZ) regime. }
\end{figure}

Next, we investigated electromagnetic surface waves at the
interface between ITO and gold in the ENZ regime. Surface plasmon
polaritons (SPPs) on metal films are receiving enormous interest
for their application in sensing and surface plasmon
optics.\cite{Barnes03} The SPP wavevector is given by
\begin{equation}\label{theta_spp}
k_{\rm SPP}=k_0 \left( \frac{\epsilon_d \epsilon_m}{\epsilon_d+\epsilon_m}\right)^{1/2} \, .
\end{equation}
The permittivities of the metal $\epsilon_m$ and the dielectric
$\epsilon_d$ both depend on frequency. The dashed black line in
Fig.~\ref{fig:dispersion}(a) shows the calculated SPP wavevector
for the ITO-gold interface. The surface plasmon resonance is
located at $0.618$~PHz, outside the plotted window. For low
energies, the SPP has a photon-like nature and its dispersion
follows that of the ITO film.

\begin{figure}[t]
\includegraphics[width=8.5cm]{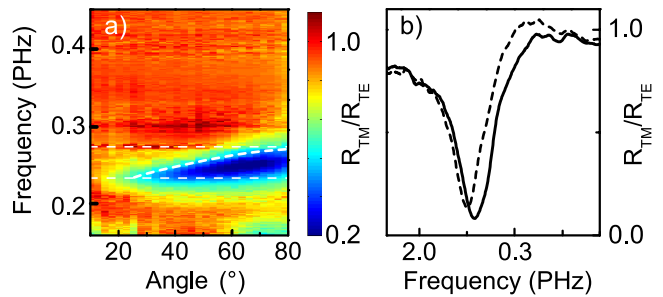}
\caption{\label{fig:itospp} (Color online) (a) Experimental
reflectivity $R_{\rm TM}$/$R_{\rm TE}$ for the ITO-Au multilayer.
(b) Spectra at $65^\circ$ angle of incidence for a 350 nm ITO - 50
nm Au (solid) and 350nm ITO - 40 nm SiO$_2$ - 50 nm Au multilayer
structures. }
\end{figure}

For conventional SPPs occurring at the interface between a
dielectric and metallic layer, the wavevector is larger than that
of free-space radiation; therefore coupling to SPPs requires
wavevector matching using a glass prism, grating, or near-field
scattering object. The ENZ-regime, however, enables a new type of
SPP mode at the interface of Au and ITO, which lies below the
light line. Consequently, the SPP can directly couple to
free-space radiation. Figure~\ref{fig:itospp}(a) shows the
measured reflectivity of the 350-nm thick ITO layer coated with a
50-nm thick (continuous) Au film, for illumination through the
ITO-side. A pronounced dip is observed around $0.25$~PHz, which
agrees well with the coupling angle obtained using $\theta_{\rm
SPP}={\rm sin}^{-1}(k_{\rm SPP}/k_0)$, which in this frequency
range closely follows $\theta_{\rm crit}$. A cross section of the
reflectivity spectrum at 65$^\circ$, plotted in
Fig~\ref{fig:itospp}(b) (solid line), shows a well-defined
spectral dip with a quality factor of $\sim 8.4$. The angular
dispersion is very flat and covers a range of around 30$^\circ$
for which light can be coupled into the SPP mode at a frequency of
0.25~PHz.

The effect of an additional SiO$_2$ spacer layer between the ITO
and Au layers is shown by the dashed line in
Fig.~\ref{fig:itospp}(b). The refractive index of the spacer
slightly increases the mode index of SPP, resulting in a redshift
of the SPP dispersion. In addition a $10\%$ narrowing is observed
which we attribute to reduced losses in the dielectric layer. The
possibility to include a spacer layer is of interest as it allows
electrically isolating the gold film from the ITO and holds
promise for incorporating active layers, or field-effect designs
for electro-plasmonic modulation.\cite{Feigenbaum10,Lu12}
Figure~\ref{fig:itospp} shows that, even for a SiO$_2$ spacer
layer of 40~nm thickness, the effective mode index of the SPP lies
below unity and the SPP couples to free-space radiation.

\begin{figure*}[t]
\includegraphics[width=15cm]{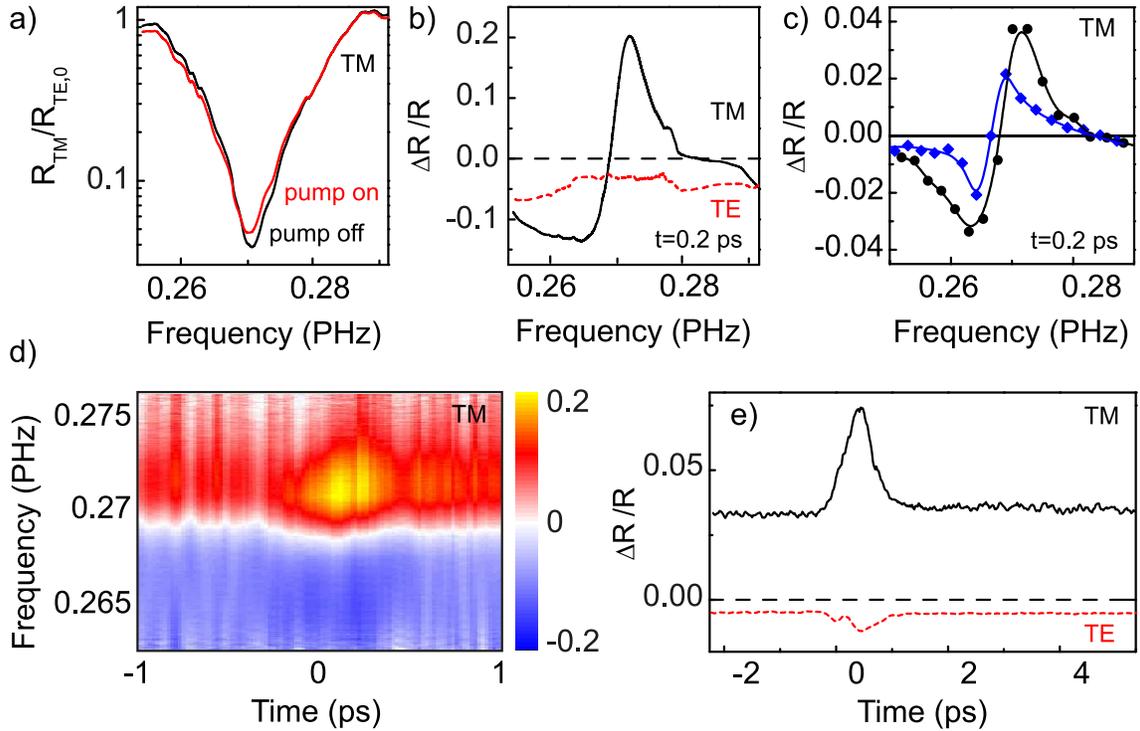}
\caption{\label{fig:pptotal} (Color online) (a) Reflectivity
$R_{\rm TM}$ with and without pump laser, normalized to intensity
of TE without pump $R_{\rm TE,0}$. (b) Differential reflectivity
$\Delta R/R$ for both TM (line, black) and TE (dash, red)
polarizations and for pump-probe delay time of 0.1~ps, at pump
fluence of 10~mJ/cm$2$. (c) Comparison of $\Delta R/R$ for ITO-Au
(dots, black) and ITO-SiO$_2$-Au (diamonds, blue) multilayers at
pump fluence of 3~mJ/cm$^2$. (d) Spectral map of differential
reflection for TM obtained using spectrometer, with (e) spectrally
integrated pump-probe signal for TM (line, black) and TE (dash,
red). (c) and (e) were measured using lock-in technique.}
\end{figure*}

As a demonstration of the functionality of the SPPs supported on
the Au-ITO interface, we demonstrate ultrafast control over the
plasmon mode using femtosecond pulsed laser excitation. Ultrafast
control of plasmons is of interest for applications in optical
switching.\cite{Macdonald08} Here we use the ultrafast response of
the Au-ITO hybrid to produce a spectral shift of the plasmon mode.
The nonlinear response was obtained using a regenerative amplifier
with OPA (Coherent RegA), producing 220-fs laser pulses with a
repetition rate of 250 kHz. Modulation of the probe pulses was
detected using a lock-in amplifier to collect the modulation
integrated over the $\sim$0.01~PHz spectral bandwidth of the OPA
output. Additionally a spectrometer could be used to resolve
relative changes in the probe spectrum larger than 1\%.
Figure~\ref{fig:pptotal}a-e show the nonlinear response of the
ITO. The absolute probe signal with and without pump is shown in
Fig.~\ref{fig:pptotal}(a) for a pump-probe delay of 0.2~ps, with
the equivalent differential reflectivity shown in
Fig.~\ref{fig:pptotal}(b) (solid line, black).\cite{remark1} For
TM-polarization, a bipolar differential reflectivity is observed,
indicative of a redshift of the SPP resonance. The magnitude of
the differential signal reaches up to 20\% at a pump fluence of
around 10~mJ/cm$2$. In comparison, the TE polarization only shows
a wavelength-independent negative contribution. The full spectral
and time evolution of the pump-probe response is shown in
Fig.~\ref{fig:pptotal}d) and e). We note, apart from an ultrafast
component, a thermal background with the same spectral shape due
to buildup of heat in the sample over many repetition cycles. The
nonlinearity is thus associated with significant energy injection,
ruling out a Kerr-type nonlinearity. The nonlinear modulation of
the SPP is therefore attributed by ultrafast heating of the Au-ITO
hybrid. Furthermore, the time-response of the TE-reflectivity
indicates the presence of an additional small instantaneous
contribution, possibly related to nondegenerate two-photon
absorption, or to the instantaneous Kerr-nonlinearity of
ITO.\cite{Elim06}

In hybrid nanostructures, hot-electron transfer may form a new
mechanism for enhancing the nonlinear
response.\cite{Abb11,Zayats12} To investigate this contribution,
we tested the response of a multilayer consisting of 50~nm Au,
20~nm SiO$_2$ and 350~nm ITO. It was found that the damage
threshold of the Au layer in this sample was substantially lower
than the Au-ITO sample by a factor of three. The reduced pump
power required the use of more sensitive lock-in detection,
resulting in the frequency dependent response shown in
Fig.~\ref{fig:pptotal}(c). We find qualitatively the same response
for the two Au-ITO (dots) and Au-SiO$_2$-ITO (diamonds, blue)
samples, indicating that in both cases, the signal is governed by
a redshift of the SPR. For the Au-SiO$_2$-ITO multilayer, the
signal is reduced by a factor two. The combination of an increased
damage threshold and higher nonlinear response in the Au-ITO
hybrid is consistent with the ultrafast hot-electron transfer
mechanism which lowers the stress of the Au-film and increases the
hybrid nonlinear response.

In conclusion, we have demonstrated new properties of ITO in the
regime characterized by a refractive index below unity (Epsilon
Near-Zero). The ITO layer shows pronounced total external
reflection which can be used to obtain waveguiding of light using
air as the high-index medium. As the refractive index and thus the
external reflection condition strongly depends on carrier density,
this effect may be useful for applications in optical sensing. In
addition, we have identified a new surface plasmon polariton mode
at the interface between ITO and gold. The plasmon polariton is
observed to couple strongly to free-space radiation without
requiring additional phase matching, which is attributed to the
strongly reduced wavevector in the ENZ regime. We have
demonstrated that this mode can be controlled on ultrafast time
scale, opening up potential application in optical switching.
Future work may identify other applications of this new plasmon
mode may be useful for applications such as outcoupling of light
from emitters, local field enhancement, or strong light absorption
in energy harvesting applications.

This work was supported by the EPSRC through grants EP/J011797/1
and EP/J016918/1.




\begin{thebibliography}{10}
\bibitem{remark1} We note that the resonance position is shifted by 0.1~PHz with respect to Fig.~\ref{fig:itospp} because
a different sample was used with slightly different carrier
density.
\end{thebibliography}
\end{document}